\documentclass[aps,preprintnumbers,floatfix,nofootinbib,preprint,superscriptaddress]{revtex4}

\usepackage{mathrsfs}
\usepackage{amsmath, amsthm, amssymb}
\usepackage{color}
\usepackage[pdftex]{graphicx}
\usepackage{amssymb}
\usepackage{verbatim}
\usepackage{hyperref}

\newcommand{\be}{\begin{equation}}
\newcommand{\ee}{\end{equation}}
\newcommand{\bea}{\begin{eqnarray}}
\newcommand{\eea}{\end{eqnarray}}

\newcommand{\df}{\dfrac}

\newcommand{\eps}{\varepsilon}
\newcommand{\fig}{Fig.}
\newcommand{\Ref}{Ref.}
\newcommand{\Eq}{Eq.}

\newcommand{\ie}{i.e.}

\begin{document}

\preprint{FT-UAM/CSIC-13-25}
\preprint{IFT-UAM/CSIC-13-097}

\title{Freeze-in through portals}
\author{Mattias Blennow}
\email{emb@kth.se}
\affiliation{Department of Theoretical Physics, School of Engineering Sciences, KTH Royal Institute of Technology, 
  Albanova University Center, 106 91 Stockholm , Sweden}

\author{Enrique Fernandez-Mart\'inez}
\email{enrique.fernandez-martinez@uam.es}
\affiliation{Departamento de F\'isica Te\'orica, Universidad Aut\'onoma de Madrid, Cantoblanco E-28049 Madrid, Spain}
\affiliation{Instituto de F\'isica Te\'orica UAM/CSIC,
 Calle Nicol\'as Cabrera 13-15, Cantoblanco E-28049 Madrid, Spain}

\author{Bryan Zald\'\i var}
\email{b.zaldivar.m@csic.es}
\affiliation{Instituto de F\'isica Te\'orica UAM/CSIC,
 Calle Nicol\'as Cabrera 13-15, Cantoblanco E-28049 Madrid, Spain}


\begin{abstract}
The popular freeze-out paradigm for Dark Matter (DM) production, relies on DM-baryon couplings of the order of the weak interactions. However, different search strategies for DM have failed to provide a conclusive evidence of such (non-gravitational) interactions, while greatly reducing the parameter space of many representative models. This motivates the study of alternative mechanisms for DM genesis. In the freeze-in framework, the DM is slowly populated from the thermal bath while never reaching equilibrium. In this work, we analyse in detail the possibility of producing a frozen-in DM via a mediator particle which acts as a portal. We give analytical estimates of different freeze-in regimes and support them with full numerical analyses, taking into account the proper distribution functions of bath particles. Finally, we constrain the parameter space of generic models by requiring agreement with DM relic abundance observations. 

 \end{abstract}

\maketitle

\section{Introduction}

By now there is solid evidence for the existence of Dark Matter (DM) from a plethora of observations such as galaxy rotation curves, structure formation, the Cosmic Microwave Background spectrum or gravitational lensing~\cite{Bertone:2004pz}. These observations comprise one of our few precious evidences of physics beyond the Standard Model. Unfortunately, all the present proofs for DM stems from its gravitational effects and thus, we remain ignorant of the particle nature of DM, \ie, its mass and interactions with the rest of the known particles, crucial ingredients so as to be able to embed it in a complete theory.
This has led to a proliferation of many different DM candidates with radically distinct phenomenologies and genesis mechanisms, with the most popular one being a Weakly Interacting Massive Particle (WIMP)~\cite{Bergstrom:2000pn}. The source of the WIMP popularity can be attributed to a combination of the Higgs hierarchy problem, whose different explored solutions typically require new weakly-interacting particles not much above the electroweak scale, and the so-called ``WIMP miracle'', referring to the fact that the correct DM relic thermal abundance can be obtained through these candidates.     

However, other mechanisms are equally viable and 
should also be explored in case DM turns out not to be a WIMP, lest we miss its signals by concentrating exclusively on the WIMP paradigm. An interesting alternative, motivated partially by the failed DM searches, is the case in which the coupling of DM to the visible sector is very suppressed. In this scenario, DM does not thermalize with the visible sector and so it tends to approach its final density from below, increasing it with increasing cross section (in contrast to the situation in the WIMP framework). This scenario has been recently referred to as freeze-in \cite{Hall:2009bx}. 

Several  mechanisms have been proposed in the past describing an out-of-equilibrium production of DM. In \cite{Dodelson:1993je} a model with sterile neutrinos as DM candidates was analysed, where these particles are populated from the thermal bath through oscillations with the SM neutrinos suppressed by small mixings. In \cite{McDonald:2001vt}, a scenario with a gauge scalar singlet DM candidate is studied, where the out-of-equilibrium DM genesis is produced from the decays of the Higgs bosons present in the thermal bath. Similar alternatives have been analysed in \cite{Yaguna:2011qn} and \cite{Frigerio:2011in}. Models where the DM candidate is produced from processes like ($\tilde b\to b \chi$), where $\tilde b$ and $b$ are thermal bath particles and $\chi$ the DM, have also been studied in the literature. An example of this has been analysed in \cite{Covi:2002vw} in the context of Supersymmetry, and afterwards in \cite{Hall:2009bx} several combinations of the masses $m_{\tilde b}$ and $m_\chi$ are analysed in detail. In \cite{DeSimone:2010tr} a case of a gravitino DM production, which is actually dominated by high temperatures has been considered~\cite{Goldberg:1983nd,Ellis:1983ew,Bolz:2000fu,Rychkov:2007uq,Cheung:2011nn}. An interesting model which is also sensitive to higher temperatures is described in \cite{Yaguna:2011ei}, where a DM candidate whose mass is larger than the reheating temperature is studied. A scenario where the portal is massless has been analysed in \cite{Chu:2011be}, while an alternative where the portal is heavier than the reheating temperature has been recently proposed in \cite{Mambrini:2013iaa}. 

In this work we will extend a possibility that was only briefly discussed in~\cite{Hall:2009bx}, namely that the weak interaction between DM and the visible sector is mediated by a portal. This is complementary to the case where the mediator is part of the thermal bath, which, as we have commented above, has been extensively analysed in the literature. The present study contains and generalises specific examples already present in the literature, such as the works presented in~\cite{Chu:2011be,Chu:2013jja,Mambrini:2013iaa, Klasen:2013ypa}. We identify the regions of parameter space where the portal (say, particle ${\cal P}$) is not in thermal equilibrium with the thermal bath, thus the scattering process ($b\bar b\to \chi\bar\chi$) can dominate over the decay process (${\cal P}\to\chi\bar\chi$) when populating the DM. We obtain analytical estimates of the predicted relic abundance coming from typical models, which we classify according to the mass of the portal. 

In section~\ref{Yield} we present the formalism used to study the evolution of the DM number density in the freeze-in regime. Section~\ref{AnalResults} is devoted to describe in detail the approximate analytical solutions of the Boltzmann equation, where the classification of different freeze-in regimes is introduced. We cross-check and fit our analytical estimates with full numerical results in section~\ref{NumResults}, and comment on the possible phenomenology of some of these models in section~\ref{pheno}, before concluding in section~\ref{concl}.

\section{DM Yield from Freeze-in}
\label{Yield}

The evolution of the DM ($\chi$-particle) number density, in the case where finite temperature effects are neglected\footnote{We refer to \cite{Hamaguchi:2011jy} for an analysis of the modified Boltzmann equation taking into account these effects.}, is described by the usual Boltzmann equation which can be expressed as:
\bea
\label{Boltzmann}
 a^{-3}\df{d (n_\chi a^3)}{dt} &=& \int\prod_{i=1}^4 \df{d^3p_i}{(2\pi)^3 2E_i} (2\pi)^4\delta^{(4)}(P_{\rm in}- P_{\rm out})\nonumber \\ 
&&\times |{\cal M}|^2 [f_b f_{\bar b}(1\pm f_\chi)(1\pm f_{\bar\chi}) - f_\chi f_{\bar\chi}(1\pm f_b)(1\pm f_{\bar b}) ]~,
\label{Boltz1}
\eea
when considering a generic 2-to-2 process $b\bar b\leftrightarrow \chi\bar\chi$ where the DM (with number density $n_\chi$) is produced from (and annihilates to) bath particle pairs $b\bar b$. The factor $a$ in the LHS is the scale factor; the index $i$ runs over the four particles, and $f_i$ are the thermal distribution functions, which for particles in thermal equilibrium will be given by $f_i = [e^{(E_i-\mu_i)/T}\mp 1]^{-1}$. Upper (lower) signs correspond to bosons (fermions).  
Furthermore, $P_{\rm in(out)}$ are the incoming (outgoing) 4-momenta of the process and $|{\cal M}|^2$ is the amplitude of the process, summed over all spins.

The freeze-in scenario assumes that the initial abundance of DM (at reheating epoch) is negligible. The thermal bath then starts populating the DM through interactions sufficiently suppressed for the DM not to thermalize with the bath. If this is the case, then the back-reaction annihilation term in \Eq~(\ref{Boltz1}), proportional to $f_\chi f_{\bar\chi}(1\pm f_b)(1\pm f_{\bar b})$, can be safely neglected, simplifying the expression to
\bea
 a^{-3}\df{d (n_\chi a^3)}{dt} &\approx&  \df{1}{8}\int_{4m_b^2}^\infty ds \int_{\sqrt{s}}^\infty dE_+ \int_{-\sqrt{E_+^2-s}}^{\sqrt{E_+^2-s}} dE_- \nonumber \\
&&\times \int\df{d^3p_\chi}{(2\pi)^3 2E_\chi}\df{d^3p_{\bar\chi}}{(2\pi)^3 2E_{\bar\chi}}\delta^{(4)}(P_{\rm in}- P_{\rm out}) 
|{\cal M}|^2 e^{-E_+/T},
\label{Boltz2}
\eea 
where we have made the approximation of $m_b\ll s$ and we have taken Maxwell-Boltzmann distribution functions ($f_i\approx e^{-E_i/T}\ll 1$). In section \ref{NumResults} we evaluate how good this approximation is when we cross-check our analytical estimates with the full numerical results which take into account the appropriate distribution functions instead. Following~\cite{Gondolo:1990dk}, the integrals over initial 3-momenta $p_b$ and $p_{\bar b}$ have been re-expressed to the variables $E_+\equiv E_b+E_{\bar b}$, $E_-\equiv E_b-E_{\bar b}$, and the centre-of-mass (squared) energy $s$. This is a convenient change of variables since the integrand in \Eq~(\ref{Boltz2}) does not depend on $E_-$. Further integration over $E_+$ and $E_-$ gives:
\bea
a^{-3}\df{d (n_\chi a^3)}{dt} \approx \df{g_b^2}{32\pi^4} \int_{4m_\chi^2}^\infty ds 
~s^{3/2} ~T~ K_1(\sqrt{s}/T) ~\sigma(s)
\eea
where $\sigma\equiv\sigma_{b\bar b\to\chi\bar\chi}\nonumber$ is the unpolarised cross-section and $g_b$ are the degrees of freedom of the $b$-particles. $K_1$ is the order-1 modified Bessel function of the second kind.

Taking into account that $a^{-3} {d (n_\chi a^3)}/{dt} = -{\bf s}HT {dY_\chi}/{dT}$, where $Y_\chi$ is the comoving number density, or yield, $(Y\equiv n/{\bf s})$, ${\bf s}$ the entropy density, and $H$ the Hubble parameter, we can finally express the DM relic density as:
\bea
\label{yield}
Y_{\chi}|_{0} =  \frac{45 M_{Pl}}{1.66 \cdot 64 \pi^6} g_b^2  \int_{T_0}^{T_{R}} dT \int_{4m_\chi^2}^\infty ds
\df{1}{\sqrt{g_*}g_*^s}\df{1}{T^5}  s^{3/2}  K_1(\sqrt{s}/T) ~\sigma(s) 
\eea
or
\bea
\label{relic}
\Omega_{\chi}h^2|_{0} &=& 2\df{m_\chi {\bf s}_0 Y_\chi|_0}{\rho_c} \\
&\approx& 3\times10^{24} ~m_\chi ~g_b^2  \int_{T_0}^{T_{R}} dT \int_{4m_\chi^2}^\infty ds \nonumber
 \df{1}{\sqrt{g_*}g_*^s}\df{1}{T^5}  s^{3/2}  K_1(\sqrt{s}/T) ~\sigma(s)
\eea
where the $0$-subindex refers to the values today, and we have considered a symmetric scenario where $n_{\bar\chi}=n_\chi$. Here $T_{R}$ is the reheating temperature, which acts as the initial condition in scenarios where the reheating epoch is assumed to be instantaneous. $M_{Pl}$ is the Planck mass and $g_* (g_*^s)$ are the energy (entropy) density effective degrees of freedom.

\begin{figure}[ht]
\centering
\includegraphics[width=0.3\textwidth,angle=0]{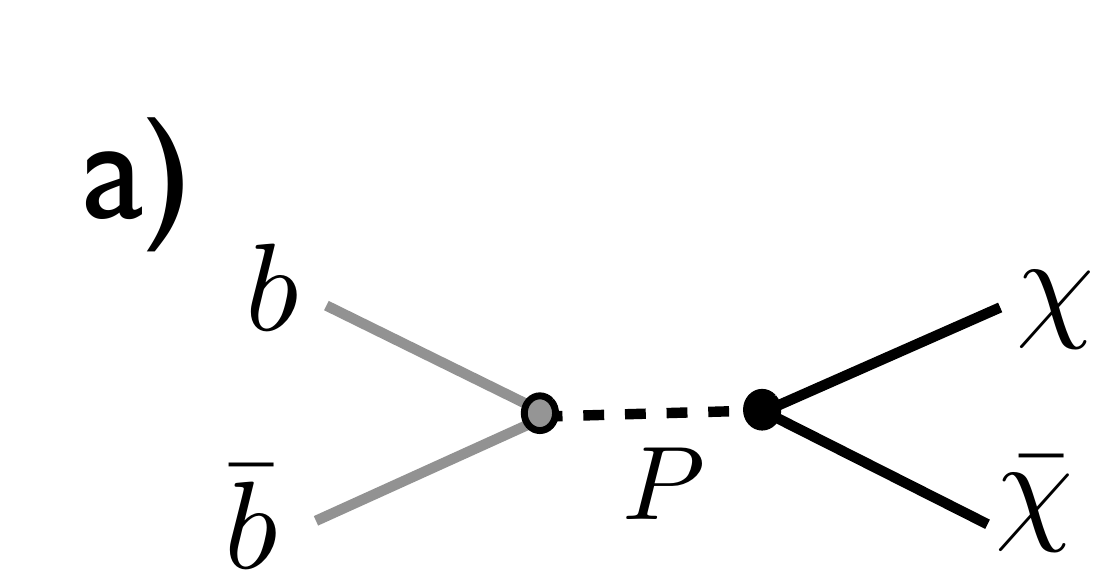}
\hspace{1cm}\includegraphics[width=0.19\textwidth,angle=0]{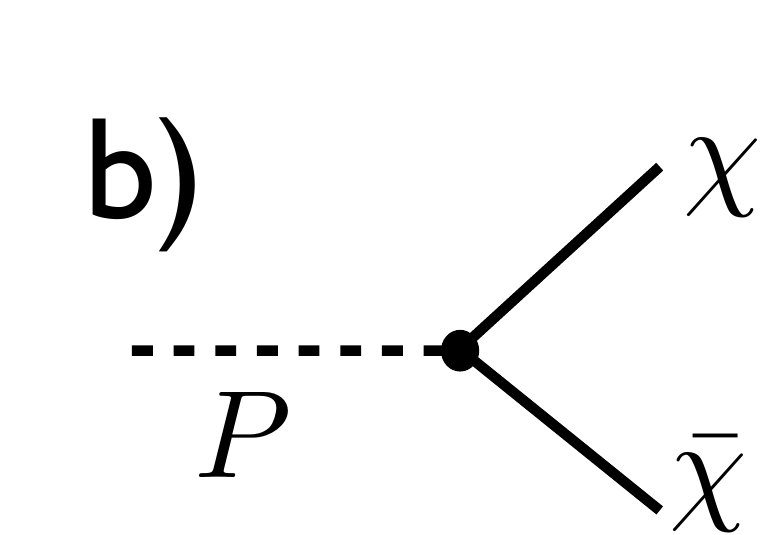}\\
\caption{\footnotesize{Feynman diagrams of processes populating the DM $\chi$. $P$ refers to the mediator (or portal), whereas $b,\bar{b}$ are bath particles.}}
\label{fig:diags}
\end{figure}

The dependence of $\sigma(s)$ on $s$ is the essential ingredient to know the behaviour of the DM yield $Y_\chi$. Here we concentrate on theories for which the DM is generated through ``portal'' interactions, where these portals are particles directly interacting with both the bath and the DM, whereas the DM only interacts directly with the portal. The dominant processes populating the DM sector are shown in \fig~\ref{fig:diags}. In what follows, we will refer to the couplings $\lambda_{BB}$ for bath-to-bath, $\lambda_{\chi P}$ for DM-to-portal interactions, and $\lambda_{BP}$ for the the bath-to-portal interactions, where either $\lambda_{BP}$ or $\lambda_{\chi P}$ should be small for the DM to be out of equilibrium.

The decay of the mediator depicted in case b) corresponds to the case that was discussed in more detail in Ref.~\cite{Hall:2009bx}. This process will tend to dominate when the portal thermalizes with the thermal bath and its mass $M$ is such that $2 m_\chi < M < T_R$. However, as we will show, there are regions of the parameter space where the portal particle does not thermalize either and production processes via freeze-in of the type depicted in a) can dominate and lead to the correct relic abundance for DM. Thus, as a complementary view to Ref.~\cite{Hall:2009bx}, in this work we will concentrate on freeze-in via a portal particle, of the type depicted in a), discussing the allowed parameter space to obtain the correct relic abundance as a function of the mediator mass, which will characterize different regimes with distinct dependence on the parameters. 
For each regime, we will derive approximate analytical expressions and compare our estimates with full numerical evaluation of the integral in \Eq~(\ref{Boltz1}), taking into account the appropriate distribution functions for the corresponding particles. 

\section{Analytical Results}
\label{AnalResults}

\begin{figure}[ht]
\centering
\input{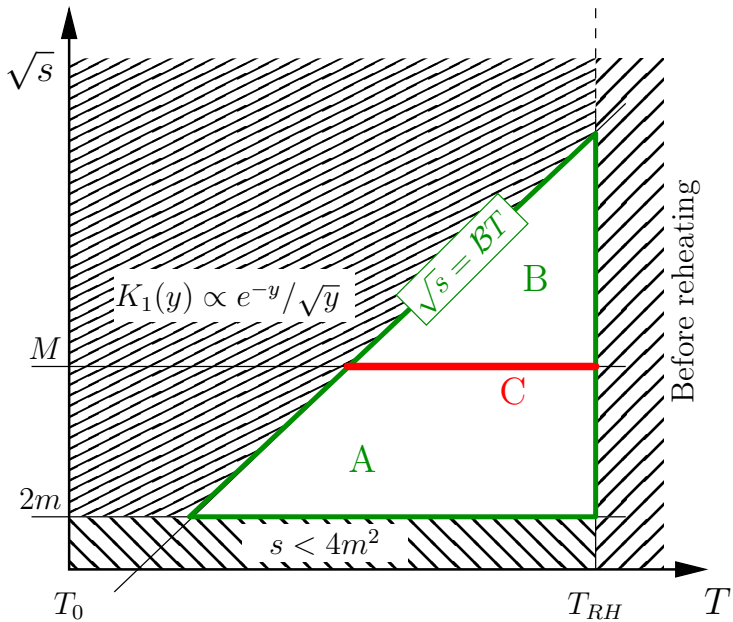}
\caption{\footnotesize{Relevant parameter space of DM production by freeze-in.}}
\label{fig:triangle}
\end{figure}

For the analytical estimates of \Eq~(\ref{yield}) it is useful to note the different limits of the Bessel function $K_1(\sqrt{s}/T)$:
\be
\label{K1}
\lim_{y\to0} K_1(y) \simeq \df{1}{y}, ~~~~ \lim_{y\to\infty} K_1(y) \simeq \df{e^{-y}}{\sqrt{y}}, ~~~~  \lim_{y\to1} K_1(y) \simeq {\cal O}(1)~.
\ee
In \fig~\ref{fig:triangle}, we depict the region of integration where the integrand is not exponentially suppressed. Essentially for $\sqrt{s}\gg T$ the DM production is negligible due to the huge Boltzmann suppression, which means that the integral over $s$ can be estimated introducing a cut-off close to $s\gtrsim T^2$. A naive estimate for the cut-off is $s^{\rm max} \simeq 9 T^2$, since $K_1(1)/K_1(3)\simeq 100$. Thus, beyond $\sqrt{s}/T = 3$ the contribution to the integral is expected to be negligible. On the other hand,  the suppression from $K_1$ may be balanced by an enhancement from the cross section $\sigma$ (see \Eq~(\ref{yield})). However, we will see below that this rough estimation is actually rather accurate. But we will keep the cut-off parameter ${\cal B}$ such that $s^{\rm max} = ({\cal B} T)^2$ free in order to compare with the exact numerical results and choose the value of ${\cal B}$ that best reproduces them, so that the analytical approximation can become an accurate proxy of the full numerical simulation.

In this section we will adopt a generic cross section for an s-channel process given by:
\be
 \sigma(s) = \frac{\lambda_{BP}^2 \lambda_{\chi P}^2}{g_b^2} \frac{s^{1/2} \sqrt{s-4m_\chi^2}}{(s-M^2)^2 + \Gamma^2 M^2}
\ee
in order to discuss the different regimes for the mediator mass $M$.


\noindent

\subsection{Heavy mediator: $M> T_R$}

We will refer to a heavy mediator of mass $M$ when $M>T_R$, where $T_R$ is the reheating temperature, which acts as a cut-off for the integral over the temperature (see \Eq~\ref{yield}). In this case, the cross-section has the following dependence with $s$:
\be
\sigma_H \approx \df{\lambda_{BP}^2 \lambda_{\chi P}^2}{g_b^2}\df{s}{M^4}~,
\label{sigmaH}
\ee
where we have assumed all the particles other than the portal to have masses $m_i^2\ll s$. As before, $\lambda_{BP}$ is the coupling of the visible (SM) sector to the portal, whereas $\lambda_{\chi P}$ is the coupling between the portal and the DM. From inspection of \Eq~(\ref{sigmaH}) it can be noted that in this case the DM production is dominated by the largest temperatures, given that the cross-section grows with $s$. We thus expect a direct dependence of the relic abundance on the reheating temperature. Indeed, for the relic abundance we obtain:

\be
\Omega_{\chi}h^2|_{0} \approx 3\times10^{24}~ m_\chi \lambda_{BP}^2 \lambda_{\chi P}^2 \left[ \df{1}{9} \df{{\cal B}^6}{g^s_*(T_R)\sqrt{g_*(T_R)}} \df{T_R^3}{M^4}\right] ~.
\label{relicH}
\ee
\noindent
This actually constitutes a special case of a more general result, where the DM production happens through an effective, non-renormalizable operator of dimension $N$ and thus suppressed by a scale $\Lambda^{4-N}$. By dimensional analysis, the relic abundance must behave as:
\be
\Omega_{\chi}h^2|_0 \propto m_\chi \df{T_R^{2N-9}}{\Lambda^{2N-8}}
\ee 
\noindent
Recently, a model with this characteristic has been presented in \cite{Mambrini:2013iaa}, which has been dubbed NETDM. The idea is that a GUT framework, e.g., $SO(10)$, can naturally provide a very heavy portal  through which the Standard Model populates the  DM. For several cases of $SO(10)$ breaking patterns, mediator masses can be larger than $10^{10}$ GeV, which require, according to \Eq~(\ref{relicH}), large reheating temperatures in order to obtain the correct relic abundance.

\subsection{Light mediator: $M< 2m_\chi$}

We define as the light regime a portal whose mass $M$ is such that $M< 2m_\chi$. In this case the cross-section can be approximated by the following expression:
\be
\sigma_L \approx \df{\lambda_{BP}^2 \lambda_{\chi P}^2}{g_b^2 s^{3/2}}\sqrt{s-4m_\chi^2}.
\label{sigmaL}
\ee
Contrary to the heavy-mediator case, now the DM production is dominated by the lowest temperatures, given the energy dependence shown in \Eq~(\ref{sigmaL}). Thus, since the DM production always stops at the ``freeze'' time ($T\lesssim 2 m_\chi$), a priori one expects that the final yield depends directly on $m_\chi$.  Indeed, the relic abundance is for this case:
\be
\Omega_{\chi}h^2|_{0} \approx 3\times10^{24}~ m_\chi \lambda_{BP}^2 \lambda_{\chi P}^2 \left[ 
\df{\pi}{12} \df{{\cal B}^3}{g^s_*(m_\chi)\sqrt{g_*(m_\chi)}}\df{1}{m_\chi}\right] ~.
\label{relicL}
\ee
Note that the $m_\chi$-dependence of the yield (proportional to the term in squared brackets above) cancels when computing the relic abundance $\Omega h^2$. It implies that for the light mediator the relic abundance is much less sensitive to the DM mass (as compared to the heavy and intermediate mediator cases). Indeed, the dependence on $m_\chi$ in this regime only stems from the effective degrees of freedom $g_* (m_\chi)$.

A particular model with these characteristics has been analysed in~\cite{Chu:2011be}, where a ``dark photon'' which acts as a massless mediator of a feeble interaction between the SM and DM is considered. 

\subsection{Intermediate mediator: $2m_\chi < M < T_R$}

There is a range of mediator masses in between the regimes described above, namely $2m_\chi < M < T_R$. In this region, the pole of the mediator propagator will be accessible during the DM production and thus we can split the integration region in three zones with qualitatively different behaviours as depicted in \fig~\ref{fig:triangle}. Within  region A, the mediator mass can be considered heavy and the integrand is given by the same expression as in the heavy mediator case, while in region B the mediator can be considered light and the integrand is given by the same expression as for the light regime. Thus, the integrals over A, and B can be approximated much in the same way as seen above. These results are typically smaller than the contribution from the peak (region C) around $\sqrt{s} \simeq M$ assuming that $\Gamma \ll M$, the Breit--Wigner peak can be approximated using the relation:
\be
 \lim_{\eps \to \infty} \frac{\eps}{\eps^2 + a^2} = \pi \delta(a),
\ee
leading to
\be
 \sigma(s) \simeq \frac{\lambda_{BP}^2 \lambda_{\chi P}^2}{g_b^2} \frac{M}{\Gamma} \pi \delta(s-M^2).
\ee
The integral over the small width of the region C is then simply taken care of by the delta function and the remaining temperature integral for the yield is
\be
 \Omega_{\chi}h^2|_0 \simeq 3\times10^{24}~ m_\chi \lambda_{BP}^2 \lambda_{\chi P}^2 \left[\frac{\pi}{3}\frac{\mathcal{B}^3}{g^s_*(M)\sqrt{g_*(M)}}\frac{1}{\Gamma}\right].
 \label{relicI}
\ee

Since in this regime the mediator is heavier than the DM particles, the mediator can decay directly into DM. If the mediator thermalizes with the bath, its decay would instead dominate the DM production through freeze-out of the mediator and its subsequent decay to DM as described in \Ref~\cite{Hall:2009bx}\footnote{The decay process ${\cal P}\to\chi\bar\chi$ is only important in this regime, since for the case of heavy mass ($M>T_R$) the mediator population is strongly Boltzmann suppressed.}. However, we will show that for large regions of the parameter space in which the intermediate mediator scenario provides the correct relic density, the mediator does not thermalize and its decay is not such an effective way of increasing the DM population.

As an example we will consider the relevant case of a vectorial mediator (e.g. a massive dark photon) coupling to the SM bath via kinetic mixing with the photon. In this scenario, the mediator will couple to SM electrons\footnote{The mediator will also couple to other charged particles but, as an example, we will show only the results from its coupling to electrons.} with $\lambda_{BP} = \lambda_0 e$, where $e$ is the electron charge and $\lambda_0$ the mixing between the mediator and the photon.  
Three processes could in principle lead to the thermalization of the mediator: its direct production through coalescence in a collision of electron and positron (notice that the cross section of this process is proportional to $\delta(\sqrt{s}-M)$); its production in $e^-$ $e^+$ annihilation in association with a photon; or via inverse Compton scattering with a dark photon instead of a photon in the final state. The rates for these processes can be found in Ref.~\cite{Redondo:2008ec}.

\begin{figure}[hbt]
\centering
\includegraphics[width=0.63\textwidth,angle=0]{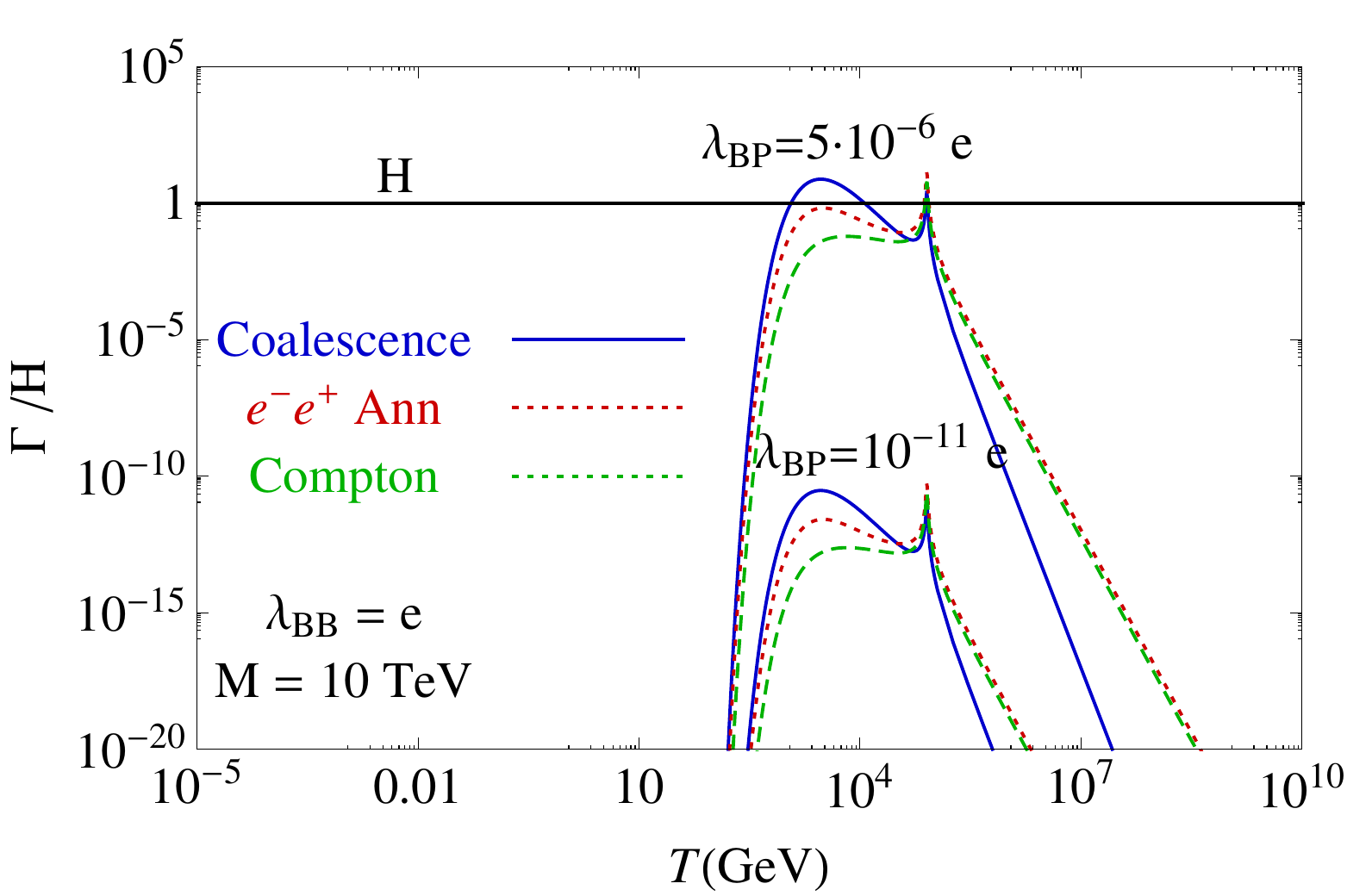}
\caption{\footnotesize{Comparison of the mediator production rates through $e^+$ $e^-$ coalescence, annihilation and inverse Compton scattering.}}
\label{fig:thermalize}
\end{figure}

In \fig~\ref{fig:thermalize} we compare these three production rates with the Hubble rate. For $\lambda_{BB}=e$, $\lambda_{BP}=10^{-11} e$ and $M=10$~TeV. We can see that all production processes are at least 10 orders of magnitude smaller than the Hubble rate for any temperature, thus preventing thermalization. As we will show with our numerical results in Sect.~\ref{NumResults}, these choices of the parameters can lead to the correct DM abundance. For $M=10$~TeV the coupling would need to be increased by about 6 orders of magnitude in order for the production rates to increase above the Hubble rate and reach thermalization. At temperatures somewhat higher than the mediator mass a spike in the production rates of the mediator appears. This spike corresponds to the temperature at which the mediator mass is equal to the thermal mass of the photon, leading to a resonantly enhanced mixing between the two~\cite{Redondo:2008ec}.

\section{Numerical Results}
\label{NumResults}

In this section we cross-check the analytical estimates made above with a full numerical analysis for the three regimes defined. For the sake of illustration, we take two type of amplitudes of DM production from the thermal bath (assumed for concreteness here to be the SM fermions), corresponding to a vector interaction: 
\be
\sigma_{\rm vec} = \df{\lambda^2_{BP}\lambda^2_{\chi P}}{32\pi^2 s^{3/2}}\sqrt{s/4-m_\chi^2} \left[2\pi \df{32}{3} \df{(s + 2 m_b^2)(s + 2 m_\chi^2)}{(s-M^2)^2 + \Gamma^2 M^2}\right]~,
\label{sigvec}
\ee
and a scalar interaction:
\be
\sigma_{\rm scal} = \df{\lambda^2_{BP}\lambda^2_{\chi P}}{32\pi^2 s^{3/2}}\sqrt{s/4-m_\chi^2} \left[ 16\pi	 \df{(s - 4 m_b^2)(s - 4 m_\chi^2)}{(s-M^2)^2 + \Gamma^2 M^2}\right]~,
\label{sigscal}
\ee
where $\Gamma$ is the total decay width of the mediator of mass $M$, taken here to be $\lambda^2_{\chi P}M/4\pi$ and $\lambda^2_{\chi P}M/8\pi$ for vector and scalar interaction, respectively. We have also assumed for simplicity only an interaction with electrons (i.e., $m_b = m_e$), but the conclusions would be similar for more complex models. 

\begin{figure}[hbt]
\centering
\includegraphics[width=0.7\textwidth,angle=0]{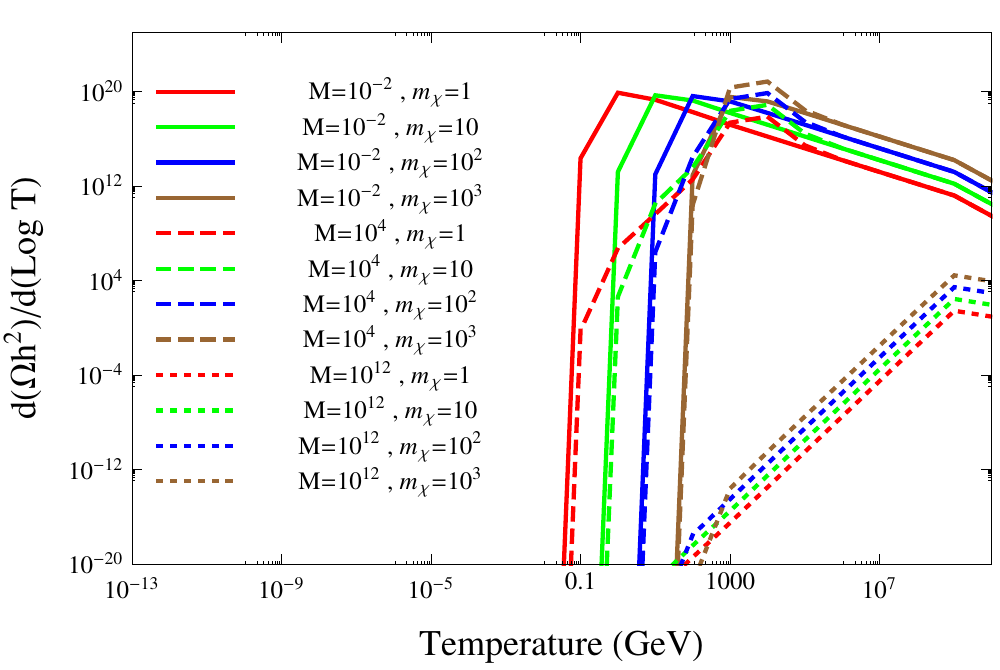}
\caption{\footnotesize{ Contribution to $\Omega h^2$ from different temperature regimes for a model with a scalar interaction between DM and the bath particles (assumed to be SM fermions). See text for more details. }}
\label{fig:scanTemp}
\end{figure}

After solving (\ref{relic}) numerically, we show an example of the contribution to $\Omega h^2$ from different temperatures in \fig~\ref{fig:scanTemp}. Just for illustration purposes, we have fixed $\lambda_{\chi P}=0.5$, and $T_R=10^{10}$ GeV. We have chosen three values of the mediator mass and several values of the DM mass. The aim of this plot is to show in which region (low or high temperatures) the production is dominantly produced. 
We have exemplified this for the model characterised by a scalar interaction giving rise to (\ref{sigscal}), but similar results apply for other types of interaction.

\noindent
For a very light mediator (in this example, $M_{M}=0.01$~GeV $\ll m_{\rm DM}^{\rm min}$) the integrand is dominated by the low temperature regime, while for the case of a very heavy mediator ($M_M=10^{12}$~GeV $\gg T_{R}$) it is the regime of high-temperatures which dominates. For an intermediate mass, we clearly observe the resonance at $T=M$, which is determined by the width of the mediator.

The numerical solutions shown in \fig~\ref{fig:scanTemp} assumed Maxwell-Boltzmann (MB) distribution functions for all particles. However, this limit is not a priori justified for freeze-in. Indeed, as shown in \fig~\ref{fig:scanTemp}, the heavy and intermediate regimes are characterized for being dominated by large temperatures, where the MB limit does not hold \footnote{Note that in the MB approximation, or when using the correct distribution functions, the factor $(1\pm f_\chi)(1\pm f_{\bar\chi})$ in \Eq~(\ref{Boltz1}) should not appear, since here we work under the assumption that DM particles are way out-of-equilibrium, which translates to $f_\chi\simeq 0$. Note also that in the more complete way of solving \Eq~(\ref{Boltz1}) would be to solve the complete integro-differential equation. However given the order of effective bath-to-DM couplings we need to obtain good relic abundances, the assumption of neglecting $f_\chi$ works extremely well.}. 


\begin{figure}[hbt]
\centering
\includegraphics[width=0.45\textwidth,angle=0]{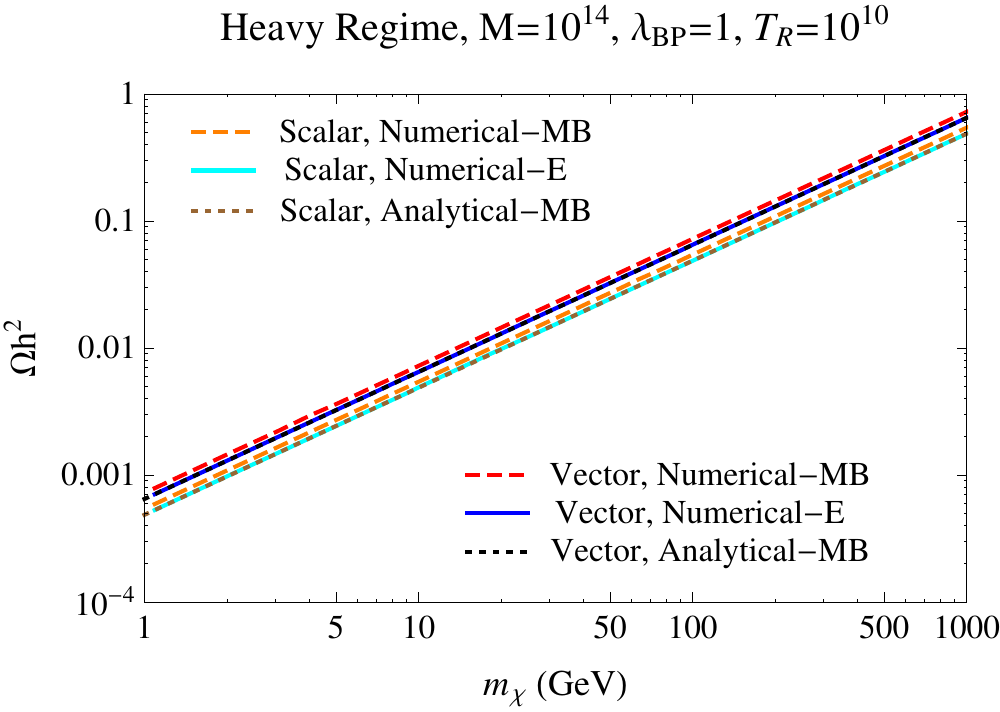}
\includegraphics[width=0.45\textwidth,angle=0]{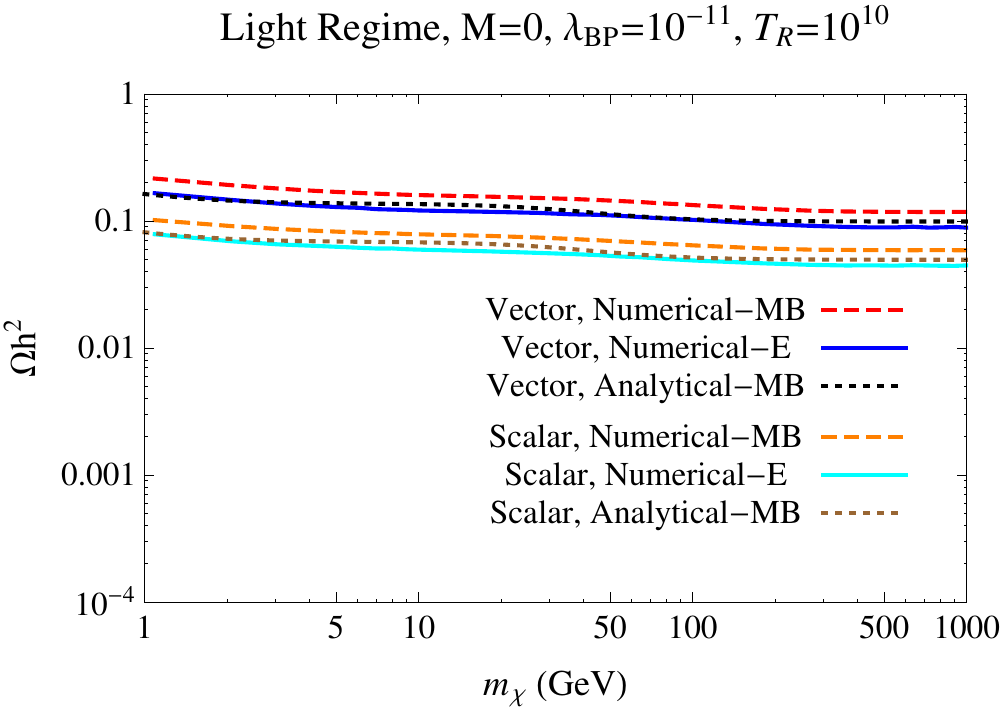}
\includegraphics[width=0.45\textwidth,angle=0]{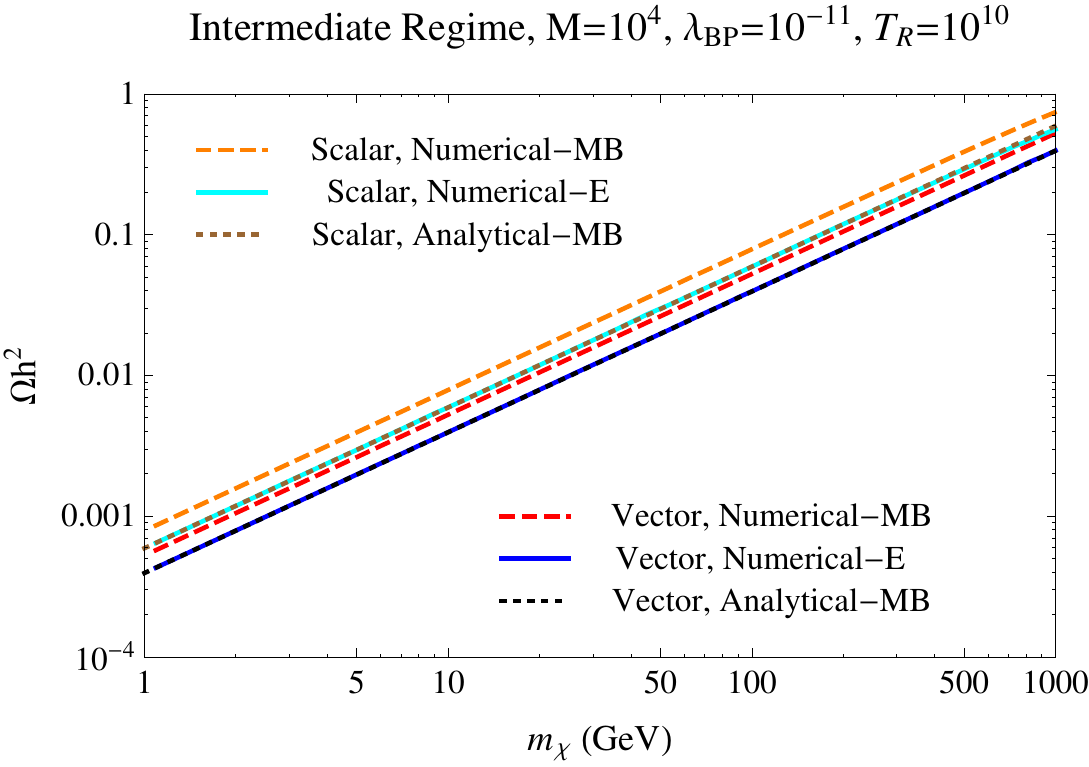}
\includegraphics[width=0.45\textwidth,angle=0]{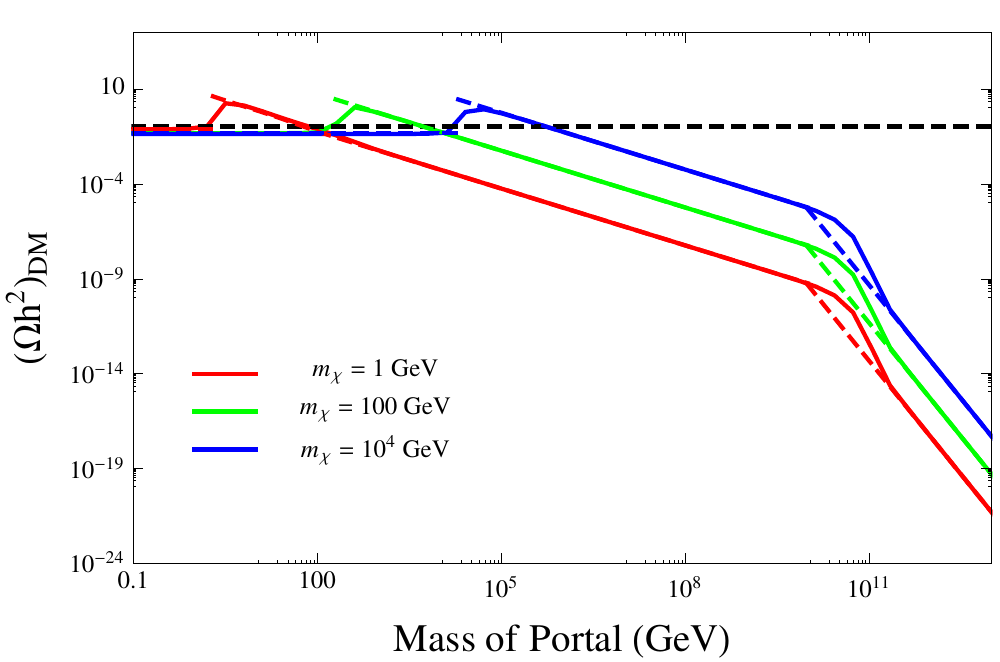}
\caption{\footnotesize{The first three panels depict the dependence of the relic density with the DM mass, for different regimes of mediator mass. Exact numerical results are shown with solid lines (blue for vector interaction and cyan for scalar interaction). Numerical results using the MB approximation are shown with dashed lines (red for vector interaction, orange for scalar one); and analytical results from Eqs.~(\ref{relicH}), (\ref{relicL}) and (\ref{relicI})  are shown with dotted lines (vector interaction in brown, whereas scalar interaction in black). The last panel compares the analytical approximations (dashed lines) for the different regimes to the exact numerical results (solid lines) as a function of the mediator mass $M$ and for different DM masses $m_\chi$.  }}
\label{fig:relic}
\end{figure} 

We compute the resulting DM abundance as a function of the DM mass in three different ways: analytically, numerically using the MB approximation, and numerically using the correct distribution functions for the bath particles. This is done for the three different regimes: heavy, light and intermediate mediator. The analytical estimation has been performed through Eqs.~(\ref{relicH}), (\ref{relicL}) and (\ref{relicI}), corrected for with appropriate factor to take into account the assumed vector (scalar) type of interaction: $1/3\pi$ ($1/4\pi$), $3/8\pi$ ($3/16\pi$) and $1/3\pi$ ($1/4\pi$), respectively. We show the results in \fig~\ref{fig:relic}. 
For the heavy regime (upper left panel) we used $M=10^{14}$ GeV for the mediator mass, $\lambda_{BP}=1$ for the bath-to-portal coupling, $\lambda_{\chi P}=1$ for the portal-to-DM coupling, and $T_R=10^{10}$ GeV for the reheating temperature. The first thing to note is a very reasonable agreement between the numerical results using both the appropriate distribution functions (solid lines) or the MB approximations (dashed lines), for the two models considered in Eqs.~(\ref{sigvec}) and (\ref{sigscal}). The discrepancy turns out to be around 15~\%. 

Regarding the analytical estimates (dotted lines), we provide in Table \ref{tab:1} the values of the ${\cal B}$-factor fitting the numerical -exact- results of each case: scalar or vector interaction, in the three different regimes.
\begin{center}
\begin{table}[hbt]
  \begin{tabular}{ | c || c | c | c | }
  \hline
  $ ~$ & Light & Intermediate & Heavy \\
    \hline\hline
    ${\cal B}$ & 2.26 & 2.17 & 3.55 \\ 
    \hline
  \end{tabular}
  \caption{Values of the ${\cal B}$-factors for the different regimes fitting the analytical with the numerical results. }
   \label{tab:1}
  \end{table}
\end{center}
We see that the fitting values are close to the naive estimation ${\cal B}\simeq 3$ made above, and they are useful to provide a very accurate analytical approximation to the exact numerical results and reproducing the correct parametric behaviour, as can be seen in \fig~\ref{fig:relic}.

In the light regime (upper right panel), we have considered a massless mediator, as well as $\lambda_{BP}=10^{-11}$, $\lambda_{\chi P}=1$ and $T_R=10^{10}$ GeV for the choice of parameters. Again, the agreement between the FD (solid lines) and MB (dashed lines) is very good and within the expectations. Concerning the analytical estimates (dotted lines), we see that they present a very good agreement with the numerical computations for ${\cal B} = 2.26$, in the expected ballpark. Besides, we note the reduced sensitivity of $\Omega h^2$ to $m_\chi$, which is encoded inside $g_*$ only.

Finally, the intermediate regime (lower left panel) is illustrated using a mediator mass of $M=10^4$~GeV, $\lambda_{BP}=10^{-11}$, $\lambda_{\chi P}=1$ and again $T_R=10^{10}$ GeV.  Note that also here the couplings $\lambda_{BP}$ should be very small in order to reproduce the value of $\Omega_{\rm DM}h^2$ measured by Planck, in a similar range to that of the light regime. 

In the final panel of \fig~\ref{fig:relic} we compare the three analytical approximations (dashed lines), including the corresponding ${\cal B}$ factors from Table \ref{tab:1}, to the numerical solution of the exact expression for the relic abundance as a function of the mediator mass $M$. We show the comparison for several choices of the DM mass and for the assumption of a scalar coupling between the portal and the fermions of the thermal bath as well as with the DM. This allows to see the transition between the three regimes defined in this work as well as the validity of the approximations. As expected, the prediction in the light regime is essentially independent of the mediator mass. In the intermediate regime, however, the relic abundance decreases linearly with $M$ (since the decay rate $\Gamma$ decreases accordingly), this is also in agreement with \Eq~(\ref{relicI}). Finally, in the heavy regime, the expected dependence with $M^{-4}$ from the heavy mediator is recovered.

\begin{figure}[hbt]
\centering
\includegraphics[width=0.7\textwidth,angle=0]{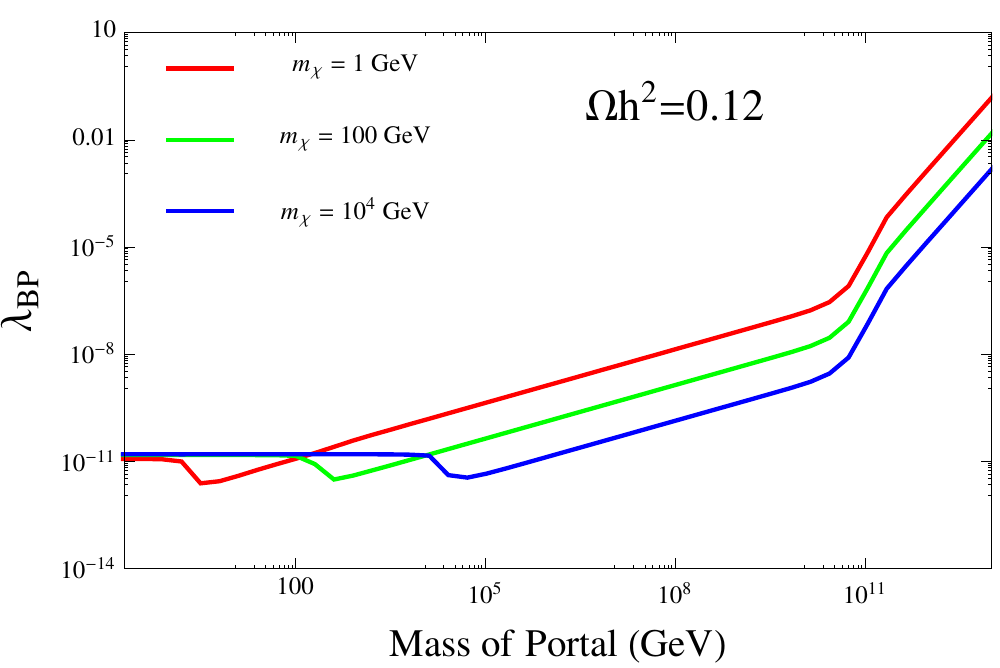}
\caption{\footnotesize{Isocontours in the portal mass $M$ and the coupling between the portal and the thermal bath $\lambda_{BP}$ plane for the correct relic density. The different contours correspond to different choices of the DM mass $m_\chi$.}}
\label{fig:bounds}
\end{figure} 
As a final result, in \fig~\ref{fig:bounds} we present the isocountours of the portal mass $M$ and the coupling between the portal and the thermal bath $\lambda_{BP}$ that lead to the correct relic density. We show this for several choices of the DM mass and for the assumption of a scalar coupling between the portal and the fermions of the thermal bath as well as with the DM. This plot exemplifies the relationships between the three parameters necessary for successful DM production and can be used for setting bounds on any of them for particular values for the others. Notice that, since in the light regime $\Omega_{\rm DM}h^2$ is independent of $M$ and very weakly dependent on $m_\chi$, a coupling $\lambda_{BP} \sim 10^{-11}$ is required to obtain the correct relic abundance.

\section{Phenomenology}
\label{pheno}

In the scenario explored here the bath-portal coupling is necessarily very small, which makes DM probes through direct, indirect and colliders searches rather challenging. However, the DM-portal coupling could be sizable. Indeed, as we have shown, the product of the two couplings should not be too small in order to obtain the correct relic abundance. A sizable DM-portal coupling would then change the paradigm of collisionless DM for structure formation, leading to DM self-interactions which could even have relatively long distance forces depending on the mass of the portal. The structure formation phenomenology of these models is thus altered, particularly at small scales, and can be probed through observations. 

On the one hand, DM self interactions can be very directly bounded by the X-ray and lensing observations of the Bullet cluster to $\sigma/m_\chi < 1$ cm$^2$/g~\cite{Randall:2007ph}, which clearly shows a separation of the luminous and dark matter components through a weaker scattering of the latter. DM self interactions can also affect the ellipticity of clusters. Indeed, self interactions would tend to thermalize the DM velocity spectrum and lead to more spherical shapes. These observations actually lead to the strongest constraints on DM self interactions $\sigma/m_\chi < 0.02$ cm$^2$/g~\cite{MiraldaEscude:2000qt}. However, these bounds have been relaxed more recently through more detailed numerical simulations which show that cross sections as large as $\sigma/m_\chi = 0.1$ cm$^2$/g~\cite{Peter:2012jh} are agreement with all observations. 

On the other hand, DM self interactions could even solve some of the experimental tensions of the standard collisionless DM simulations.  Indeed, self-interactions can mediate energy transfer from the outer halo to the central region leading to softer profiles for dwarf galaxies, alleviating the cuspines characteristic of collisionless DM simulations that is not in good agreement with observations~\cite{Spergel:1999mh}. Similarly, if DM is not collisionless, dwarf subhaloes could be stripped via collisions depleting the abundance of Milky Way satellites, which simulations with collisionless DM tend to overproduce. 

In Ref.~\cite{Tulin:2013teo} the authors compute the velocity-dependent transfer cross section as a function of the masses of DM, the portal and the coupling between the two. With the average of the cross section over the relevant velocity scales they derive approximate bounds on this parameter space, as well as estimate the preferred regions to alleviate the collisionless DM paradigm shortcomings. They conclude that, if $m_\chi \leq 100$~GeV, then the mass of the portal $M>100$~MeV ($M>10$~MeV) for $\lambda_{\chi P} \sim 1$ ($\lambda_{\chi P} \sim 0.1$). A wide range of the parameter space with $m_\chi \leq 1$~TeV, $\lambda_{\chi P} \sim 0.1$ and $M \sim$ few MeV seems to alleviate the shortcomings of collisionless DM while being in agreement with present bounds. For larger DM masses $m_\chi$, data can still be accommodated for smaller masses of the portal $M$. 
This phenomenology is complementary to that of the scenario with a more strongly coupled portal in the thermal bath and very feebly interacting dark matter, which would not lead to these modifications of structure formation but would typically present more prominent phenomenology at colliders~\cite{Hall:2009bx}.

Nevertheless, the light regime can also be probed in direct detection experiments; specifically for mediators lighter than the recoil energy $E_r$, since the scattering cross-section has an infrared divergence as $E_r^{-2}$, even if the coupling is very tiny. An interesting alternative analysing electron recoils has been presented in \cite{Essig:2012yx}, but more common experimental studies based on nuclear recoils have promising prospects, being able to test sufficiently feeble couplings in a few years from now (see e.g. \cite{Chu:2011be}).   

Finally, for the heavy regime the phenomenology is much more challenging given that the main phenomenology of these models takes place at the reheating period, for which there are at present no direct probes. Apart from the allowed range $1{\rm MeV} \lesssim T_R \lesssim 10^{14}{\rm GeV}$, the lower bound coming from BBN and the upper bound from a typical prediction of chaotic inflationary models \cite{Linde:2005ht}, there is no prospect for constraining $T_R$ better than this in the near future. Assuming that the observed DM content comes solely from a heavy-mediated candidate, the above range can be translated into 
\be
9\times10^3 m_\chi^{1/4} \lesssim ~\left(\df{M}{1{\rm GeV}}\right) \lesssim 10^{17} m_\chi^{1/4}
~.
\ee

\section{Conclusions}
\label{concl}

In this work we have concentrated in the so-called freeze-in mechanism for dark matter (DM) production. In this framework, the genesis of DM happens out of thermal equilibrium, since its connection to the thermal bath is assumed to be very suppressed. Thus, if there is a portal mediating this interaction, the product of the couplings of the portal with the bath and DM must consequently be small. Here we have focused on scenarios in which the portal is also out of thermal equilibrium because the bath-to-portal coupling is suppressed, while the DM-to-portal coupling could be sizable. This scenario is complementary to the more discussed freeze-in case where DM genesis occurs from the out-of-equilibrium decays of a particle which is part of the thermal bath, characterized by a sizable bath-to-portal coupling, whereas DM is only feebly interacting. 

We have performed analytical estimates of the DM relic abundance for the different regimes that can be identified according to the mass of the portal. These analytical results are based on the assumption that the distribution function of all particles follow a Maxwell-Boltzmann law, which a priori is not justified in processes for which the temperature $T$ is much greater than the masses of all relevant particles during production. Furthermore, the resulting Bessel function has been approximated by a simplified expression in the region of interest. 
We have studied the size of the corrections driven by these simplifications by cross-checking our analytical estimates with a complete numerical analysis. We found that the Maxwell-Boltzmann approximation is reasonable with a discrepancy of around $15\%$. On the other hand, the analytical approximation, while maintaining the correct parametric behaviour, strongly depends on how the production region is approximated. By comparing with the exact numerical results, we have obtained the size of the integration region which allows to reproduce with very good accuracy these results with the simple analytical expressions derived. Thus the analytical expressions can be safely used instead of the exact numerical results with the corresponding correcting factor. Finally, we have used the exact numeric results to set constraints on the parameter space of generic models (i.e. masses and couplings of DM and the portal) so as to obtain a correct DM abundance as measured by WMAP9~\cite{Hinshaw:2012aka} or more recently, Planck~\cite{Ade:2013zuv}.  

\begin{acknowledgments}
We thank Y. Mambrini and J. Redondo for very useful discussions. This work was supported by the G\"oran Gustafsson Foundation (M.B.). E.F.M acknowledges financial support by the European Union through the FP7 Marie Curie Actions CIG NeuProbes (PCIG11-GA-2012-321582) and the ITN INVISIBLES (PITN-GA-2011-289442); the Spanish MINECO through the Ram\'on y Cajal programme (RYC2011-07710) and through the project FPA2009-09017; and the Comunidad Aut\'onoma de Madrid
through the project HEPHACOS P-ESP-00346.
B.Z. acknowledges the Consolider-Ingenio PAU CSD2007-00060, CPAN CSD2007-00042, under the contract FPA2010-17747.
E.F.M. and B.Z. acknowledge the hospitality of KTH Royal Institute of Technology during the completion of this work as well as the support from the G\"oran Gustafsson Foundation making this visit possible.
\end{acknowledgments}

\bibliography{bfmz}{}

\end{document}